\documentclass{article}

\usepackage[intlimits]{amsmath}
\usepackage{cite}
\usepackage{epsfig}

\renewcommand{\d}{\partial}

\renewcommand{\k}{{\bf k}}
\newcommand{\p}{{\bf p}}
\newcommand{\T}{\textrm{T}}

\newcommand{\rh}{\varrho}
\newcommand{\ep}{\varepsilon}

\newcommand{\exv}[1]{\left\langle{\,#1\,}\right\rangle}

\newcommand{\Tr}{\mathop{\textrm{Tr}}}
\newcommand{\sgn}{\mathop{\textrm{sgn}}}

\renewcommand{\Im}{\,\textrm{Im}\,}
\renewcommand{\Re}{\,\textrm{Re}\,}
\newcommand{\pint}[2]{{\int\!\frac{d^{#1}#2}{(2\pi)^#1}\,}}

\renewcommand{\c}[1]{{\cal{#1}}}

\def\tr{{\rm Tr\,}}

\def\re{{\rm Re\,}}

\def\bea{\begin{eqnarray}}
\def\eea{\end{eqnarray}}
\def\nn{\nonumber}
\def\half{\frac{1}{2}}

\def\d{\partial}

\def\lmatrix{\left(\begin{array}}
\def\rmatrix{\end{array}\right)}

\newcommand{\ZZ}{\Delta}

\begin{document}

\vspace{-1cm}

\hfill UCSD/PTH 08-08

\vspace{2cm}

\begin{center}{\Large\bf Shear viscosity of pure Yang-Mills theory at strong coupling}\\
\vspace{0.5cm}
Antal Jakov\'ac$^{\,a}$ and D\'aniel N\'ogr\'adi$^{\,b}$\\
\vspace{0.5cm}

$^a${\em Department of Theoretical Physics, University of Wuppertal\\ Gau$\beta$strasse 20, Wuppertal 42119, Germany }

\vspace{0.5cm}

$^b${\em Department of Physics, University of California, San Diego\\ 9500 Gilman Drive, La Jolla, CA 92093, USA }
\end{center}

\begin{abstract}
  We calculate the shear viscosity to entropy density ratio in pure $SU(N)$
  Yang-Mills theory below the critical temperature using the strong coupling expansion. 
  The result for the $\eta/s$ ratio for
  temperatures around the phase transition is $0.22$ for $N = 2$ and $0.028N^2$ for $N > 2$.
  The results are 
  consistent with the conjectured $1/4\pi$ lower bound inspired by the AdS/CFT correspondence.
\end{abstract}

\section{Introduction}

Ideas originating from the AdS/CFT correspondence found a surprisingly rich set of applications
in the description of strongly coupled QCD plasma. At first sight this may indeed
seem surprising since the prototype case of ${\cal N} = 4$ supersymmetric Yang-Mills
theory is very different from QCD. For instance the former is a conformal theory
while the latter is not. However, the application is to QCD in the deconfined phase
and for large temperatures QCD is actually close to being conformal. There are other 
reasons too why it is perhaps not an accident that AdS/CFT inspired ideas work
better than expected \cite{Shuryak:2003xe}.

One particular instance where AdS/CFT based arguments provide a good description
of experimental data is the case of transport coefficients.
In heavy ion experiments a fairly good description of the data can be achieved
using (classical) hydrodynamical models and the coefficients of the
corresponding Navier-Stokes equation can be determined from measurements.
These and in particular the $v_2$ value of the anisotropy of the observed
flow \cite{Adler:2003kt, Adams:2003am} point to the fact that the QCD plasma
is an almost perfect fluid. 
A dimensionless quantity can be formed from another coefficient,
the shear viscosity $\eta$ and the entropy density $s$ which experimentally
turns out to be rather small, $\eta / s \approx 0.2$ \cite{Teaney:2003kp, Shuryak:2003xe}.

Based on the aforementioned AdS/CFT ideas a universal lower bound was conjectured
$\eta / s \geq 1 /4\pi$ for a large class of quantum field theories 
\cite{Policastro:2001yc, Kovtun:2004de, Son:2007vk}. This bound is known to 
be saturated by ${\cal N } = 4$ 
super Yang-Mills theory at large 't Hooft coupling and large $N$ in which limit
the gravity dual is weakly coupled and the calculation is tractable on the gravity side. 
The first corrections at finite $N$ and 't Hooft coupling was shown to be positive and
hence comply with the conjectured bound \cite{Buchel:2004di, Benincasa:2005qc, Buchel:2008sh,Myers:2008yi}.

One argument supporting the universal lower bound conjecture is that the energy-time uncertainty
relation can be reformulated entirely in terms of $\eta/s$ using the quasi particle picture \cite{Kovtun:2004de}.
The lower bound of the uncertainty relation yields a lower bound for $\eta/s$.

However the precise set of quantum field theories where the bound is expected to hold
is not clearly formulated. Counter examples are known which served to limit the conjectured
validity of the bound \cite{Cohen:2007qr, Son:2007xw, Cohen:2007zc}. As a result it is not clear what precise 
statement should be proved or disproved and a case-by-case approach seems more fruitful \cite{Kats:2007mq, Brigante:2007nu, Brigante:2008gz, Ge:2008ni}.
Some of the studied cases satisfy the bound while some do not. 

The counter examples in \cite{Cohen:2007qr, Cherman:2007fj} fall into two types.
In one of them the number of particle species is very large, in
the other the number of excitations of a particle are very large. While these
constructions are convincing perhaps the conjectured bound can be valid in a limited sense.
In any case this issue is far from being settled \cite{Son:2007xw, Cohen:2007zc}.

The second type of construction in \cite{Cherman:2007fj} may be realised in the framework of quantum
field theory as infinitely many peaks finally melting into a continuum. This
suggests that a good candidate for violating the proposed lower bound 
should contain a strong continuum preferably with no quasi particle
peaks in the spectral function i.e. it should be neither Bose nor Fermi gas.

Apart from models with known gravity duals analytic computations were done in QCD 
using the quasi particle approach
\cite{Arnold:2000dr, Arnold:2003zc, Arnold:2006fz} and recently using
perturbation theory in QED \cite{Gagnon:2007qt}. These studies yield
$\eta/s \sim 1/g^4\ln g^{-1}$. Clearly, when perturbation theory is applicable
the coupling is small and the above ratio is large. And vica versa, a small
measured value of this quantity hints at a large value of the coupling
constant.  Newer calculations \cite{Xu:2007ns, Xu:2007jv}
suggest however that higher order quasi particle processes may still lead to small
viscosity even in the perturbative regime.

A non-perturbative calculation of $\eta/s$ has been attempted using a lattice
discretization in $SU(3)$ pure gauge theory \cite{Karsch:1986cq, Meyer:2007ic, Meyer:2008dq}.
Unfortunately even in cases where a reasonable ansatz is available for the spectral function e.g. a 
transport peak structure, quantitative details are very difficult to extract from
correlation functions \cite{Petreczky:2007js}.

Using the large $N_f$ expansion of the 2PI effective action, which formalism sums an infinite
set of diagrams, transport coefficients were calculated in \cite{Aarts:2005vc}. It was found that for
large fermion masses the shear viscosity can be arbitrarily small.

In this paper the $\eta / s$ ratio of pure $SU(N)$ Yang-Mills theory
will be calculated on a Euclidean lattice at strong coupling for $T < T_c$.
The actual final result is $0.22$ for $N = 2$ and $0.028N^2$ for $N > 2$ at
temperatures close to the deconfinement transition. 
We find a vanishing $\eta/s$ ratio at $T\to0$, however the systematic uncertainties 
inherent in the strong coupling expansion are hard to estimate.
Similar calculations
for transport coefficients have been carried out in \cite{Horsley:1985dz,Horsley:1985fr}.

The strong coupling expansion has several conceptual difficulties not specific to
the observable we calculate \cite{Montvay:1994cy}. Since on the lattice with
decreasing lattice spacing the bare coupling constant is going to zero, we
cannot approach the continuum limit. The commonly
used strategy is trying to express the results in terms of physical quantities,
which are then interpreted as renormalized quantities. Another issue is the
presence of the roughening transition which makes it impossible to reach
temperatures higher than a certain value.
Despite of these difficulties, however, the strong coupling expansion not only works rather
well for describing several concepts in the confined phase such as chiral symmetry breaking
and confinement but also quantitatively \cite{Montvay:1994cy}.

The outline of the paper is as follows. First the shear viscosity is calculated
in section \ref{sec:shear}. The calculation is based on the Kubo formula,
which will be related to the lattice plaquette-plaquette correlator. In
the strong coupling regime the plaquette-plaquette correlators are dominated
by elongated tube-like surfaces. A subclass of such simple surfaces
as well as more complicated ones involving a loop wrapping around the finite
imaginary time direction are resummed in section \ref{sec:T1loop} leading
to a final formula for $\eta$. In section \ref{sec:entropy} the entropy is calculated 
from the
derivative of the free energy. In section \ref{sec:ratio} we calculate the
shear viscosity to entropy ratio in the limit of the temperature being close
to the transition point. Section \ref{sec:discussion} concludes with a discussion
of our findings.

\section{Shear viscosity}
\label{sec:shear}

The shear viscosity is one of the linear response coefficients coming from the
response of the energy momentum tensor to an external current coupled to the
energy momentum tensor itself \cite{Hosoya:1983id}. Given the correlation
function
\begin{equation}
\label{uuu}
  C(x) = \frac{1}{10} \exv{[\pi_{ij}(x),\pi_{ij}(0)]},
\end{equation}
where $\pi_{ij} = T_{ij} - \delta_{ij} T_{kk} / 3$ is the trace-less part of
the spatial components of the energy momentum tensor, the shear viscosity is
the leading coefficient of its Fourier transform $C(\omega,\k)$ for small frequency and
zero spatial momentum,
\begin{equation}
  \label{visc}
  \eta = \lim\limits_{\omega\to0} \frac{C(\omega,\k=0) }{\omega}\,.
\end{equation}

First we assume that we are at zero temperature and find a simple
representation for the energy-momentum correlation function. We will work in
Euclidean space and start with the definition,
\begin{equation}
  M_{\mu\nu\,\rh\sigma}(x) = \exv{\T\left( T_{\mu\nu}(x) T_{\rh\sigma}(0)
    \right)}\,,
\end{equation}
from which the following symmetry properties are evident,
\begin{equation}
\label{sym}
  M_{\mu\nu\,\rh\sigma} = M_{\nu\mu\,\rh\sigma} = M_{\mu\nu\,\sigma\rh} =
  M_{\rh\sigma\,\mu\nu}\,.
\end{equation}

For the shear viscosity we will compute the following Euclidean expression:
\begin{equation}
\label{blabla}
  \c M(x) =\frac1{10}\sum\limits_{ij}\exv{\T \left( \pi_{ij}(x)\pi_{ij}(0) \right) }\,.
\end{equation}

At zero temperature we have $SO(4)$ rotational invariance in Euclidean
space which can be used to relate the value of $\c M$ at a generic
$x_\mu$ argument to the value of $M_{\mu\nu\,\rh\sigma}$ at $\tau=(|x|,0,0,0)$.

\subsection{Symmetries}

To this end $M_{\mu\nu\,\rh\sigma}$ is decomposed into invariant tensors compatible
with the symmetry properties (\ref{sym}).
The most general such tensor depending on $x_\mu$ can be written as, 
\begin{eqnarray}
  \label{inv1}
  M_{\mu\nu\,\rho\sigma}(x) = && M_1(|x|)\, \delta_{\mu\nu}\delta_{\rh\sigma}
  + M_2(|x|) \left(\delta_{\mu\rh}\delta_{\nu\sigma} +
    \delta_{\nu\rh}\delta_{\mu\sigma}\right) + M_3(|x|) \frac{ x_\mu x_\nu
    \delta_{\rho\sigma} + x_\rho x_\sigma \delta_{\mu\nu}
  }{ x^2 } + \nn \\
  && M_4(|x|) \frac{ x_\mu x_\sigma \delta_{\nu\rh} + x_\nu x_\sigma \delta_{\mu\rh}
    + x_\mu x_\rho \delta_{\nu\sigma} + x_\nu x_\rho \delta_{\mu\sigma} }{x^2} +
  M_5(|x|)\, \frac{ x_\mu x_\nu x_\rho x_\sigma }{ x^4 } \,, 
\end{eqnarray}
where the coefficient functions $M_1$, $M_2$, $M_3$, $M_4$ and $M_5$ can only
depend on $|x|$. Using this formula at coordinate $\tau = (|x|, 0, 0, 0 )$ we
obtain the following special cases
\begin{eqnarray}
  M_{00\,00}(\tau) &=& M_1(|x|) + 2 M_2(|x|) + 2 M_3(|x|) + 4 M_4(|x|) + M_5(|x|) \nn \\
  M_{00\,11}(\tau) &=& M_1(|x|) + M_3(|x|) \nn \\
  M_{01\,01}(\tau) &=& M_2(|x|) + M_4(|x|) \\
  M_{11\,11}(\tau) &=& M_1(|x| )+ 2 M_2(|x|) \nn \\
  M_{11\,22}(\tau) &=& M_1(|x|)\,. \nn 
\end{eqnarray}
From the above formulae one can express the functions $M_{1,2,3,4,5}$ as linear combinations
of certain components of $M_{\mu\nu\,\rho\sigma}$, 
\begin{eqnarray}
  \label{inv2}
  M_1(|x|) &=& M_{11\,22}(\tau)\nn \\
  M_2(|x|) &=& \half M_{11\,11}(\tau) - \half M_{11\,22}(\tau) \nn \\
  M_3(|x|) &=& M_{00\,11}(\tau) - M_{11\,22}(\tau) \\
  M_4(|x|) &=& M_{01\,01}(\tau) - \half M_{11\,11}(\tau) + \half M_{11\,22}(\tau) \nn \\
  M_5(|x|) &=& M_{00\,00}(\tau) + M_{11\,11}(\tau) -2M_{00\,11}(\tau)
  -4M_{01\,01}(\tau). \nn 
\end{eqnarray}
After contracting $M_{\mu\nu\,\rho\sigma}$ appropriately in order to get $\c M(x)$ at any
generic $x_\mu$ argument we arrive at,
\begin{equation}
  \c M(x)= M_2(|x|) +\frac{2}3 u M_4(|x|) + \frac{2}{30} u^2 M_5(|x|),
\end{equation}
where $u=x_i^2/x_\mu^2$. This can be rewritten using (\ref{inv2}) as
\begin{eqnarray}
  \c M(x) &=& \frac{2}{30} u^2 M_{00\,00}(\tau) - \frac{4}{30} u^2
  M_{00\,11}(\tau) + \frac{u}{15}(10-4u)M_{01\,01}(\tau) + \nn \\ 
  && + \frac{1}{30}(15-10u+2u^2)M_{11\,11}(\tau) + \frac{1}{6}(2u-3)M_{11\,22}(\tau)\,.
\end{eqnarray}
In the above formula the $M_{\mu\nu\,\rho\sigma}$ components on the right hand
side are evaluated at the point $\tau=(|x|,0,0,0)$.

\subsection{Plaquette-correlators}

All of the above manipulations were done in the continuum and now we will switch
to the lattice by writing some of the energy momentum tensor components in
terms of plaquette variables $P_{\mu\nu}(x)$ \cite{Karsch:1986cq},
\begin{equation}
  T_{\mu\mu} = \frac{\beta}{N} \left[ - \sum\limits_{\nu\not=\mu} P_{\mu\nu}
    + \sum_{\sigma,\nu\not=\mu; \sigma>\nu} P_{\sigma\nu} \right]\,,
\end{equation}
or more explicitly,
\begin{eqnarray}
  && T_{00} = \frac{\beta}{N} \left[ -P_{01} -P_{02}-P_{03} +P_{21} +P_{31} +P_{32} \right] \nn \\
  && T_{11} = \frac{\beta}{N} \left[ -P_{10} -P_{12}-P_{13} +P_{20} +P_{30} +P_{32} \right] \nn \\
  && T_{22} = \frac{\beta}{N} \left[ -P_{20} -P_{21}-P_{23} +P_{10} +P_{30} +P_{31} \right] \\
  && T_{33} = \frac{\beta}{N} \left[ -P_{30} -P_{31}-P_{32} +P_{20} +P_{10} +P_{21} \right]\,. \nn
\end{eqnarray}
Substituting the above back into $\c M$ we find,
\begin{eqnarray}
\label{hejho}
  \c M(x) =&&  \frac{2}{15} \frac{\beta^2}{N^2} (15 - 10u - 4u^2 ) \biggl(2P_{10\,21}(\tau)
    -2P_{10\,32}(\tau) + P_{10\,10}(\tau) + P_{21\,21}(\tau) -\nn\\ 
    &&-P_{10\,20}(\tau) - P_{21\,31}(\tau)\biggr) + \frac{4}{30} u(5u-2)M_{01\,01}(\tau).
\end{eqnarray}
where $P_{\mu\nu\,\rh\sigma}(x) = \exv{P_{\mu\nu}(x)P_{\rh\sigma}(0)}$ is the
plaquette-plaquette correlator.

As we will see, at small $\beta$ i.e. large coupling, the leading and equal contributions come from
$P_{10\,21}$ and $P_{21\,21}$. Hence we can write
\begin{equation}
  \c M(x) = \frac{2}{5} \frac{\beta^2}{N^2} (15-10u-4u^2)\, P_{21\,21}(\tau)\,,
\end{equation}
at leading order.

For the shear viscosity we need the Fourier transform of $\c M$ at
momentum $p=(p_0,0,0,0)$ where $p_0\to 0$,
\begin{equation}
  \c M(p) = \frac{2}{5} \frac{\beta^2}{N^2}  \int\! d^4x\, (15-10u-4u^2)
  e^{i p_0 t}  P_{21\,21}(\tau).
\end{equation}
The $P_{21\,21}$ term only depends on $|x|$ and the terms containing the $u$
variable only depend on the angles. For small $p_0$ one may substitute the
$15 - 10u - 4u^2$ term with its integral over the angles, 
\begin{equation}
  \label{MandP}
   \c M(p_0,\p=0) = \frac{2\beta^2}{N^2}\,\int\!d^4 x\, e^{i p_0 t}
   P_{21\,21}(\tau)\,.
\end{equation}
Thus the plaquette-plaquette correlator is needed for plaquettes in the $21$ plane.

\subsection{Strong coupling expansion}
\label{csakany}

The plaquette-plaquette correlator, $G_{pp}(\tau)$, will be calculated via
the strong coupling expansion.

The Wilson lattice formulation of pure $SU(N)$ gauge theory is used and the action is,
\begin{equation}
  \label{wilsonaction}
  S = \sum_p S_p = \beta \sum_p \left( 1 - \frac{1}{N} \re \tr U_p \right)\,,
\end{equation}
where $\beta = 2N/g_0^2$, $g_0$ is the bare coupling constant, the sum is over
all plaquettes $p$ of the lattice and $U_p$ is the plaquette variable built
out of the links $U_\mu(x)$ in the usual way.

The strong coupling expansion amounts to an expansion in small $\beta$; for
details see \cite{Montvay:1994cy} or \cite{Smit:2002ug}. Using the characters
$\chi_R$ of $SU(N)$ there is an expansion which sums over all irreducible
representations $R$,
\begin{equation}
  \label{exp}
  e^{ - S_p } = c(\beta) \left( 1 + \sum_{R\neq 0} d_R\, a_R(\beta)\, \chi_R(
    U_p ) \right)\,,
\end{equation}
where $d_R$ is the dimension of the representation and $c(\beta)$ and
$a_R(\beta)$ are some coefficients. These coefficients will be expanded in
$\beta$.

At leading order for the group $SU(N)$ we have \cite{Smit:2002ug},
\bea
\label{eq:af}
  c(\beta) &=& 1 \nn \\
  a_f(\beta) &=& \left\{ \begin{array}{l}
      \frac{\beta}{N^2}\;\, \qquad N=2 \\ \\
      \frac{\beta}{2N^2} \qquad N> 2 \end{array} \right.\,.
\eea
where $f$ refers to the fundamental representation.

The plaquette-plaquette correlator can be computed from the partition function
if the coupling $\beta$ is changed temporarily to $\beta_1$ and $\beta_2$ at
the two plaquettes and remains $\beta$ elsewhere
\begin{equation}
\label{abc}
  \exv{P_1(\tau)P_2(0)} = N^2 \frac{\ln Z(\beta_1,\beta_2,\beta)}
  {\d\beta_1\d\beta_2}\biggr|_{\beta_1=\beta_2=\beta}\,.
\end{equation}

The logarithm of the partition function is computed by forming closed
connected surfaces and assigning representations to the plaquettes of the
surface following the usual recipe of the strong coupling expansion
\cite{Smit:2002ug}. To leading order the fundamental representation is assigned
to each plaquette leading to the contribution 
\begin{equation}
  d_f^2 a_f^n(\beta),
\end{equation}
where $a_f(\beta)$ is the expansion coefficient from (\ref{exp}) corresponding
to the fundamental representation, $d_f = N$ is its dimension and $n$ is the
number of plaquettes on the surface. The leading order terms in the expansion
of $a_f(\beta)$ were given in (\ref{eq:af}).

The smallest surfaces spanning between well separated plaquettes are tube-like
objects: they are closed at both end, and go almost straight from one
plaquette to the other. An extra factor comes from the fact that the
orientation of the tube and the orientation of the plaquette is not
necessarily the same. In table \ref{tab:factors} we summarize these extra
factors for the orientations that appear in (\ref{hejho}).
\begin{table}[htbp]
  \centering
  \begin{tabular}[c]{|c|c|c|c|c|c|c|}
    \hline
 $P_{10\,21}$ & $P_{10\,32}$ & $P_{10\,10}$ & $P_{21\,21}$ & $P_{10\,20}$ &
 $P_{21\,31}$ & $M_{01\,01}$ \cr
 \hline
 1 & $a_f^4$ & $a_f^4$ & 1 & $a_f^4$ & $a_f^4$ & $<a_f^2$\cr
    \hline
  \end{tabular}
  \caption{Extra factors necessary for closing the surface in case of
    different orientations of the ending plaquettes. The tube is in the
    temporal direction.} 
  \label{tab:factors}
\end{table}
To estimate the value of $M_{01\,01}$ we recall that in terms of plaquettes its expression
is \cite{Karsch:1986cq},
\begin{equation}
  M_{01\,01}(x) = \frac{4\beta^2}{N^2}\sum\limits_{\sigma,\rh} \exv{ \left(\Tr \tilde
      P_{0\sigma}(x) \tilde P_{1\sigma}(x) \right) \left(\Tr \tilde
      P_{0\rh}(0) \tilde P_{1\rh}(0) \right) },
\end{equation}
where
\begin{equation}
  \tilde P_{\mu\nu}(x) = -\frac i2\left(
    P_{\mu\nu}(x)-P_{\mu\nu}(x)\right)_\mathrm{traceless\,part}.
\end{equation}
The minimal surface which contains all the plaquettes of the above formula contains
(in addition to the tube contribution) extra 2 plaquettes. Therefore the 
4-plaquette expectation value
has an extra contribution of $a_f^2$ which explains the last column of table \ref{tab:factors}.
Hence we are indeed left with the single contribution of $P_{21\,21}$ as promised.

The logarithm of the partition function in \eqref{abc} at leading order is then
\begin{equation}
  \ln Z(\beta_1,\beta_2,\beta) = d_f^2 a_f(\beta_1) a_f(\beta_2) G^E(\tau)\,,
\end{equation}
where $G^E(\tau)$ is the open tube contribution, leading to
\begin{equation}
  P_{21,21}(\tau) = \c A_N G^E(\tau)\,,
\end{equation}
where
\begin{equation}
  \c A_N = N^4 \frac{a_f^2(\beta)}{\beta^2} =  \left\{
    \begin{array}[c]{ll}
      1\qquad & \mathrm{for}\; N=2 \cr
      1/4\qquad & \mathrm{for}\; N>2 \cr
    \end{array}
    \right.
\end{equation}

Therefore the quantity \eqref{MandP} we are interested in is simply,
\begin{equation}
  \label{MandGlat}
   \c M(p_0,\p=0) = \frac{2 \beta^2 \c A_N}{N^2} \, G^E(p_0,\p=0)\,.
\end{equation}

\subsection{Straight line contribution}

Now we proceed with the calculation of the open tube correlation function $G^E(\tau)$.

The lowest order contribution from the surface which includes plaquette $p$ on
the hypersurface at $t=0$ and at $t=\tau$, both having the same spatial position, is the straight tube giving,
\begin{equation}
  \label{GppandG}
   G_\mathrm{str}^E(\tau) = e^{-\sigma\tau}\,,
\end{equation}
where $\sigma = - \frac{4}{a} \ln a_f(\beta)$ is the glueball mass. The
subscript ``str'' stands for straight. For generic end points we have
$G_\mathrm{str}^E(x) = \exp(-\sigma|x|)$.

We will need its Fourier transform,
\begin{equation}
  \label{hejho1}
  G_\mathrm{str}^E(k) = \frac1{a^4}\int d^4x e^{-ikx -\sigma |x|} = 
  \frac \ZZ {(k^2/\sigma^2+1)^{5/2}}\,,
\end{equation} 
where
\begin{equation}
\label{delta}
  \ZZ = \frac{12\pi^2}{(a\sigma)^4}.
\end{equation}

To evaluate the shear viscosity we have to calculate the imaginary part of the
retarded propagator of the corresponding energy-momentum tensor components,
and up to now we calculated the Euclidean propagator. But the retarded and the
Euclidean propagators are related as
\begin{equation}
\label{hejho2}
  G^E(-i\omega + \ep,\k) = iG^\mathrm{ret}(\omega,\k)\,,
\end{equation}
whereas from the retarded Green function one obtains the spectral function as
\begin{equation}
  \rh(k_0,\k) = \mathop{\mathrm{Disc}}_{k_0} iG^\mathrm{ret}(k_0,\k)
   = -2\Im iG^\mathrm{ret}(k_0+i\ep,\k)\,,
\end{equation}
where Disc stands for discontinuity. Thus the retarded Green function from
(\ref{hejho1}) and (\ref{hejho2}) is
\begin{equation}
  iG_\mathrm{str}^\mathrm{ret}(k) = \frac \ZZ{(1-k^2/\sigma^2)^{5/2}},
\end{equation}
with Landau prescription, $k_0\to k_0+i\ep$. The spectral function then
reads,
\begin{equation}
  \rh_\mathrm{str}(k) = \mathop{\mathrm{sgn}}(k_0) \Theta(k^2-\sigma^2)
  \frac{2\ZZ}{(k^2/\sigma^2-1)^{5/2}}\,.
\end{equation}
The spectral function exhibits a mass gap; it is zero below $\sigma$,
the glueball mass.

\subsection{Zig-zagging lines}

Now that we have the contribution of a straight line (tube) we will also add
the contribution of zig-zagging lines with arbitrary number of joints.

\begin{figure}[htbp]
\label{fig:zigzag}
\centering
\includegraphics[height=3cm]{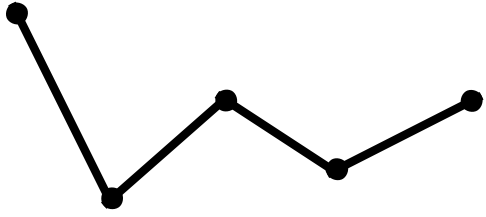}
\caption{Diagrams consisting of an arbitrary number of joints. Each line is a
  thin tube.}
\end{figure}

\begin{figure}[htbp]
\label{fig:spectralfunction}
\centering
\includegraphics[height=7cm]{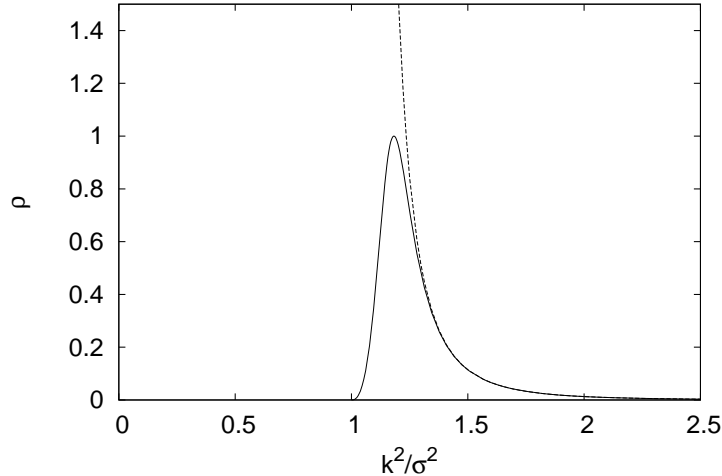}
\caption{The spectral function at $\Delta=0.1$ (solid line) and unresummed
  (dashed line). The two curves differ only in the threshold region. The
  resummed spectral function exhibits a peak and a smooth threshold behaviour.}
\end{figure}

A zig-zagging line with one joint has the following contribution,
\begin{equation}
  \sum_z G_\mathrm{str}^E(z) G_\mathrm{str}^E(x-z) =
  \frac1{a^4}\int\! d^4z\,
  G_\mathrm{str}^E(z)G_\mathrm{str}^E(x-z)\,, 
\end{equation}
which turns into $G_\mathrm{str}^E(k)^2$ in Fourier space. Similarly, the
contribution of zig-zagging lines with $n$ joints is
$G_\mathrm{str}^E(k)^{n+1}$. Summing them all leads to,
\begin{equation}
  G_\mathrm{str}^E(k) + {G_\mathrm{str}^E(k)}^2 +
  {G_\mathrm{str}^E(k)}^3 + \cdots =
  \frac1{{G_\mathrm{str}^E(k)}^{-1} - 1}\,.
\end{equation}
The calculation is completely analogous to the usual self-energy summation in
ordinary perturbation theory. The spectral function corresponding to this
resummation is 
\begin{equation}
  \label{rhoform}
  \rh_0(k) = \mathop{\mathrm{sgn}}(k_0) \Theta(k^2-\sigma^2) \bar\rh (k)\,, 
\end{equation}
where we have introduced
\begin{equation}
\label{spec}
  \bar\rh(k) = \frac{2\ZZ (k^2/\sigma^2-1)^{5/2}}{\ZZ^2+(k^2/\sigma^2-1)^5}\,.
\end{equation}
The spectral function is plotted on Fig.~\ref{fig:spectralfunction}.

By resumming the contribution of all zig-zagging lines with an arbitrary
number of joints and arbitrary location for these joints we have essentially
resummed the contribution of all curves since any curve can be approximated by
zig-zagging lines. This resummation changed the spectral function in an
essential way only near the threshold: it goes to zero at the threshold at
$k^2=\sigma^2$ and exhibits a peak at
\begin{equation}
  \label{kmax}
  k_\mathrm{max}^2=\sigma^2(1+\ZZ^{2/5}),\qquad \rh_\mathrm{max} = 1.
\end{equation}
Note that it smeared out the divergence at the threshold of the straight tube
case. Nonetheless the zig-zagging resummation really affects only the vicinity
of the threshold; for example, at $k^2=\sigma^2(1+2\ZZ^{2/5})$ the zig-zagging
propagator is smaller than the straight line propagator only by $3\%$.

The spectral function above may be interpreted as a quasi particle (with mass
$M=\sigma\sqrt{1+\ZZ^{2/5}}$, cf. \eqref{kmax}, and width proportional to
$\sigma\ZZ^{1/5}$). We must, however, not forget about the presence of the
continuum, which may alter some of the consequences coming from the
quasi particle assumption.

\subsection{1-loop contribution and finite temperature}
\label{sec:T1loop}

So far we were at zero temperature where according to \eqref{spec} the
contribution to the viscosity is zero because of the presence of the mass gap.
At finite temperature the finite imaginary time direction and periodic boundary
conditions has to be taken into
account leading to,
\begin{equation}
  G^E_T(\tau,x) = \sum\limits_{n=-\infty}^\infty G^E_0(\tau+n/T,x)\,.
\end{equation}
This, however, does not change the spectral function and only introduces the usual Bose factor
$n(\omega) = (\exp(\omega/T) - 1 )^{-1}$ into the spectral representation of
the Green functions.
The zero temperature spectral representation,
\begin{equation}
  iG^E_0(\tau) = a\int\limits_0^\infty\!\frac{d\omega}{2\pi}
  e^{-\omega|\tau|} \rh_0(\omega),
\end{equation}
turns into
\begin{equation}
  \sum\limits_{n=-\infty}^\infty G^E_0(\tau+n/T) =
  a\int\limits_0^\infty\!\frac{d\omega}{2\pi} \left( e^{\omega(1/T-|\tau|)} +
    e^{\omega|\tau|}\right) n(\omega) \rh_0(\omega)\,,
\end{equation}
at finite temperature where on the right hand side the zero temperature
spectral function appears. As a consequence the
retarded Green function will remain the same as at zero temperature,
$G^\mathrm{ret}_T = G^\mathrm{ret}_0$.

So far we have included all single tube contributions, even the ones wrapping around
the imaginary time direction, but did not find any nonzero contribution
below the mass gap $\sigma$. This is expected also when we examine local
modifications of the propagator, since these only modify the value of the
mass gap \cite{Montvay:1994cy,Langelage:2008dj}. As a hint for what types
of diagrams could be relevant, we recall that at weak coupling the phenomenon
of Landau damping is similar; it yields nonzero contribution to the imaginary
part of the propagator below the light cone at finite temperature, while this
region is excluded at zero temperature. The most important
diagram resulting in Landau damping is a loop diagram, where one propagator is
wrapping around the imaginary axis, while the other is not. 

\begin{figure}
  \centering
  \includegraphics[height=3cm]{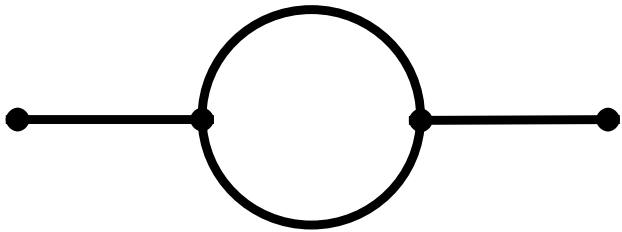}
  \caption{The one loop diagram contribution to the viscosity. Each line is a
    thin tube.} 
  \label{fig:loop}
\end{figure}

We also include the effect of a 1-loop strong coupling diagram, see figure \ref{fig:loop}.
In complete analogy to ordinary perturbation theory a self-energy can be introduced
and the retarded self-energy can be expressed from
the zero temperature one as
\cite{weakcouplingpertth}
\begin{equation}
  \label{Sigmaret}
  \Sigma^\mathrm{ret}(p) = \frac{a^4}{a_f^2(\beta)} \pint4k \left(\frac12 +
    n(k_0)\right) \rh_0(k) G^\mathrm{ret}_0(p-k)\,.
\end{equation}
since each of the two joints of the one-loop diagram gives a factor of $a_f(\beta)^{-1}$.

Using this retarded self-energy the full one-loop corrected spectral
function $\rh_\mathrm{1-loop}(p)$ corresponding to the plaquette-plaquette
correlator is
\begin{eqnarray}
  \rh_\mathrm{1-loop}(p)&=& \mathop{\mathrm{Disc}}_{p_0}
  iG^\mathrm{ret}_\mathrm{1-loop}(p) = 
  \mathop{\mathrm{Disc}}_{p_0} \frac1{(iG_0)^{-1}(p) - \Sigma^\mathrm{ret}(p)}
  = \nn \\ &=& \frac{-\mathop{\mathrm{Disc}}_{p_0} \left( (iG_0)^{-1}(p) -
      \Sigma^\mathrm{ret}(p) \right)} {(\Re (iG_0)^{-1}(p) - \Re
    \Sigma^\mathrm{ret}(p))^2 + (\Im (iG_0)^{-1}(p) - \Im
    \Sigma^\mathrm{ret}(p) )^2}\,.
\end{eqnarray}
For the viscosity the above expression is needed at zero spatial momentum and small
frequency where $G_0^{-1}$ has no imaginary part, only a non-vanishing real part. On the other
hand $\Sigma^\mathrm{ret}(p)$ can be neglected to leading order in the denominator. Therefore
near zero frequency we have,
\begin{equation}
  \rh_\mathrm{1-loop}(p_0\approx 0,\p=0) \approx
  \frac{\mathop{\mathrm{Disc}}_{p_0} \Sigma^\mathrm{ret}(p)} {
    (\Re(iG_0)^{-1}(0))^2} = (iG_0(0))^2 
  \mathop{\mathrm{Disc}}_{p_0}\Sigma^\mathrm{ret}(p).
\end{equation}
Since $iG^{ret}_0(k=0) = G^E_0(k=0)$ we obtain,
\begin{equation}
  \label{rhot}
   \rh_\mathrm{1-loop}(p_0\approx 0,\p=0) = \ZZ^2
   \mathop{\mathrm{Disc}}_{p_0}\Sigma^\mathrm{ret}(p).
\end{equation}
From \eqref{Sigmaret} we find
\begin{equation}
  \label{haliho}
  \mathop{\mathrm{Disc}}_{p_0}\Sigma^\mathrm{ret}(p)  = \frac{N^4
    a^4}{\beta^2\c A_N} \pint4k \left(\frac12 + n(k_0)\right) \rh_0(k)
  \rh_0(p-k),
\end{equation}
where we also used \eqref{eq:af} to substitute the leading order value for
$a_f(\beta)$. 

Using \eqref{rhoform} at $\p = 0$ the following four combinations of step functions
will appear in (\ref{haliho}),
\begin{eqnarray}
  && \Theta(k_0-\omega_\k)\Theta(p_0-k_0-\omega_\k)\,, \nn \\
  && \Theta(-k_0-\omega_\k)\Theta(-p_0+k_0-\omega_\k)\,,\nn \\
  && -\Theta(k_0-\omega_\k)\Theta(-p_0+k_0-\omega_\k)\,,\\
  && -\Theta(-k_0-\omega_\k)\Theta(p_0-k_0-\omega_\k)\,. \nn
\end{eqnarray}
The first two is nonzero only when $p_0\ge 2\omega_\k$ or $p_0\le
-2\omega_\k$, respectively. Therefore they do not give contribution at small
frequency. The remaining two terms yield for $0\le p_0 \le\sigma$,
\begin{equation}
   \mathop{\mathrm{Disc}}_{p_0}\Sigma^\mathrm{ret}(p_0,\p=0)  = -\frac{N^4
     a^4}{\beta^2\c A_N} \pint4k \left(n(k_0) - n(k_0-p_0)\right)
   \Theta(k_0-p_0-\omega_\k) \bar\rh(k) \bar\rh(k-p),
\end{equation}
Clearly for vanishing momentum
$\mathop{\mathrm{Disc}}_{p_0}\Sigma(0)=0$ as it should be. The linear term in
$p_0$ is then,
\begin{equation}
  \lim_{p_0\to0}
  \frac{\mathop{\mathrm{Disc}}_{p_0}\Sigma^\mathrm{ret}(p_0)}{p_0}  =
  \frac{N^4 a^4}{\beta^2\c A_N} \pint4k 
  \left(-\frac{d n(k_0)}{dk_0} \right) \Theta(k_0-\omega_\k) \bar\rh^2(k).
\end{equation}

A further simplification is possible since it is consistent with all previous
approximations to truncate the expansion
\begin{equation}
  -\frac{dn(k_0)}{dk_0} = \frac1T\,\frac{e^{k_0/T}}{(e^{k_0/T}-1)^2} =
  \frac1T\,e^{-k_0/T} + \c O(e^{-2k_0/T}).
\end{equation}
at the first term because the neglected terms contribute as much as higher loop
diagrams. This leads to,
\begin{equation}
  \lim_{p_0\to0}
  \frac{\mathop{\mathrm{Disc}}_{p_0}\Sigma^\mathrm{ret}(p_0)}{p_0}  =
  \frac{N^4 a^4}{4\pi^3T\beta^2 \c A_N} \int\limits_0^\infty dk
  k^2\int\limits_{\omega_\k}^\infty dk_0 e^{-k_0/T} \bar\rh^2(k)\,.
\end{equation}
which in turn yields the following result for the spectral function corresponding
to the plaquette-plaquette correlator using (\ref{rhot}),
\begin{equation}
  \frac{\rh_\mathrm{1-loop}(p_0\approx 0,\p=0)}{p_0} = \frac{\ZZ^2 N^4
    a^4}{4\pi^3T\beta^2 \c A_N} \int\limits_0^\infty dk k^2
  \int\limits_{\omega_\k}^\infty dk_0 e^{-k_0/T} \bar\rh^2(k)\,.
\end{equation}

Ultimately we can express $\eta$ collecting all relations from (\ref{uuu}),
(\ref{visc}), (\ref{blabla}) and (\ref{MandGlat})
\begin{equation}
\label{viscfinal}
  \eta = \frac{\ZZ^2 N^2 a^4}{2\pi^3T} \int\limits_0^\infty
  dk k^2 \int\limits_{\omega_\k}^\infty dk_0 e^{-k_0/T} \bar\rh^2(k)\,.
\end{equation}

\section{The entropy}
\label{sec:entropy}

The entropy density is the derivative of the free energy density with respect
to temperature.

The free energy in the strong coupling expansion is coming from closed and
connected surfaces similarly to the discussion in section (\ref{csakany}).
The first temperature dependent contribution to the free energy and hence
the first non-vanishing contribution to the entropy comes from surfaces wrapping
around the imaginary time direction.

The first contribution is again from a thin straight tube line wrapping around
once. The value of the dimensionless free energy density corresponding to
such a straight tube has been calculated in \cite{Langelage:2008dj}
\begin{equation}
  f = -\frac{3}{\sqrt{\c A_N}} (aT) a_f(\beta)^{4N_t}\,,
\end{equation}
where $1/\sqrt{\c A_2} = 1$ and $1/\sqrt{\c A_{N>2}} = 2$ comes from the fact that for $N > 2$ the
fundamental representation and its complex conjugate are two different representations.
The factor of 3 comes from the 3 possible orientation of the tube (12, 13 and
23).

In exactly the same way as was done for the calculation of the viscosity one needs
to take zig-zagging as well as 1-loop diagrams into account.
To be able to collect these effects the free energy density is written as
\begin{equation}
  f = -\frac{3}{\sqrt{\c A_N}} (aT) \left( G_T^E(0) - G_{T=0}^E(0) \right)\,.
\end{equation}
Now we can use the spectral representation of the propagator and write
\begin{equation}
  f = - \frac{3}{\sqrt{\c A_N}} Ta^5 \pint4k \left(\frac12+n(k_0)-\frac12\sgn(k_0)\right)
  \rh(k)\,.
\end{equation}
The approximation $n(k_0>0)\approx e^{-k_0/T}$ is valid just as before since the higher 
loop diagrams are not taken into account. Within this approximation one arrives at,
\begin{equation}
  f = -\frac{3}{\sqrt{\c A_N}} Ta^5 \pint4k e^{-|k_0|/T} \rh(k)\,,
\end{equation}
and using the specific form of the spectral function from \eqref{rhoform} we
find
\begin{equation}
  f = - \frac{3Ta^5}{2\pi^3 \sqrt{\c A_N}}  \int\limits_0^\infty \!dk
  k^2\int\limits_{\omega_\k}^\infty \!dk_0\, e^{-k_0/T} \bar\rh(k)\,.
\end{equation}

The entropy density from here is then,
\begin{equation}
\label{entropyfinal}
  s = \frac{3a^5}{2\pi^3 \sqrt{\c A_N}}  \int\limits_0^\infty \!dk
  k^2\int\limits_{\omega_\k}^\infty \!dk_0\, \left( 1 + \frac{k_0}T\right)
  e^{-k_0/T} \bar\rh(k) \approx \frac{3a^5}{2\pi^3 T\sqrt{\c A_N}}
  \int\limits_0^\infty \!dk k^2\int\limits_{\omega_\k}^\infty \!dk_0\, k_0\,
  e^{-k_0/T} \bar\rh(k)\,.
\end{equation}
Note that the resulting entropy is independent of $N^2$
\cite{Langelage:2008dj,Csernai:2006zz}. This result will be used in the next
section to calculate the shear viscosity to entropy density ratio.

\section{The shear viscosity to entropy density ratio}
\label{sec:ratio}

Our final goal is the ratio $\eta/s$ and using (\ref{viscfinal}) and (\ref{entropyfinal}) from
the previous sections one obtains,
\begin{equation}
  \label{etaperes0}
  \frac\eta s = \frac{N^2\Delta^2\sqrt{\c A_N}}{3} \; \frac{ \exv{ \bar\rh^2(k) }_T} {
    \exv{a k_0 \bar\rh(k)}_T }\,, 
\end{equation}
where the following short hand notation is introduced 
\begin{equation}
  \exv{F}_T = \pint4k \Theta(k_0) \Theta(k^2-\sigma^2)\, e^{-k_0/T}
  \, F(k)\,.
\end{equation}
It is worth emphasizing that all dependence on $N$ have dropped out completely
(even for finite $N$) as expected from naive counting of degrees of freedom. Both the entropy density and
the shear viscosity are expected to scale as $N^2$ and hence the ratio is expected
to be independent of $N$ in accordance with our formula (\ref{etaperes0}).

In order to evaluate (\ref{etaperes0}) expressions of the type $\exv{(ak_0)^n \bar\rh^m(k)}_T$
are needed. The computation of these terms is greatly simplified
by the fact that in the strong coupling expansion $T/\sigma$ is small. 

\begin{figure}[htbp]
\label{fig:etaperes}
\centering
\includegraphics[height=7cm]{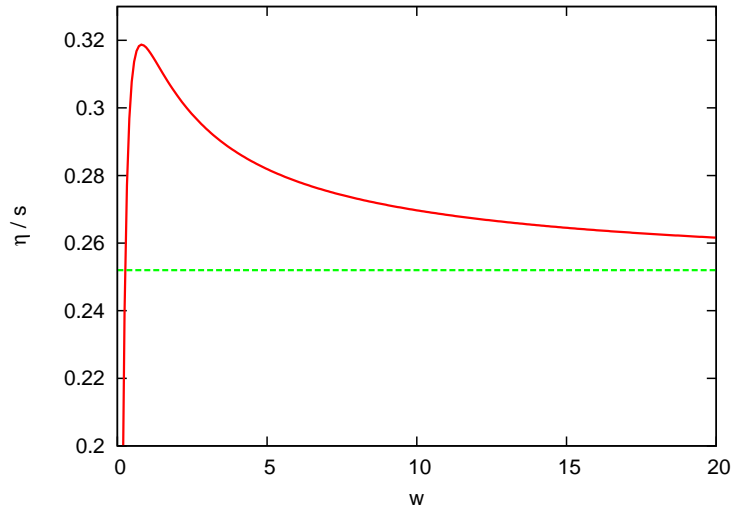}
\caption{The $\eta/s$ ratio as a function of the rescaled temperature $w$ for
  $\Delta = 1$ and $N = 3$. The horizontal line represents the limiting value $0.252$ which is
  taken as the $T \to T_c$ estimate.}
\end{figure}

A convenient coordinate system to use is
\begin{equation}
  k_0 = k \cosh \eta,\qquad k_x = k\sinh\eta\sin\theta\cos\phi,\qquad k_y =
  k\sinh\eta\sin\theta\sin\phi,\qquad k_z = k\sinh\eta\cos\theta\,,
\end{equation}
and the relevant quantities then are given by
\bea
\exv{(ak_0)^n \bar\rh^m(k)}_T = \frac{1}{4\pi^3} \int_\sigma^\infty dk k^3 (ak)^n \, \bar\rh^m(k) \int_0^\infty d\eta\, \sinh^2\eta\,
\cosh^n\eta\, e^{-k\cosh\eta/T}\,.
\eea
A lengthy but straightforward manipulation of the integral in the
approximation that $T/\sigma$ is small leads to
\bea
\label{nmfinal}
\exv{(ak_0)^n \bar\rh^m(k)}_T = \sqrt{\frac{\pi}{2}} \frac{T^{5/2} \sigma^{3/2}}{4\pi^3} e^{\sigma/T} (a\sigma)^n \, w^{-1}
\int_0^\infty dz\, e^{-z/w}\, \left( \frac{2z^{5/2}}{1+z^5} \right)^m\,.
\eea
where $w = 2\, \ZZ^{-2/5}\, T / \sigma$
and (\ref{spec}) has been used for the spectral function $\bar\rh(k)$.
Note that all $n,m$-independent factors in (\ref{nmfinal}) will drop out from
the ratio $\eta/s$, which finally can be written as
\begin{equation}
  \label{etaperes_sce}
  \frac\eta s = N^2 \sqrt{\c A_N} \ZZ^{9/4} f( \frac{2T}{\ZZ^{2/5}\sigma}), \qquad
  \mathrm{where}\qquad f(w) = 
  \frac1{3(12\pi^2)^{1/4}} \frac{\displaystyle \int_0^\infty dz\, e^{-z/w}\,
      \left( \frac{2z^{5/2}}{1+z^5} \right)^2} {\displaystyle \int_0^\infty
      dz\, e^{-z/w}\, \left( \frac{2z^{5/2}}{1+z^5} \right)}.
\end{equation}
This is the first main result of our paper. The plot of $f(w)$ -- which is the
$\eta/s$ ratio in the $\ZZ=1$ limit -- can be seen on Fig. \ref{fig:etaperes}.

We can further analyze the expression $f(w)$ in the small and large $w$ regions.
In the large $w$ limit we can omit the exponential, and we obtain the
largest contribution from the peak of the spectral function. As a result we
obtain a $w$-independent form. In the small $w$ regime, on the other hand, the
exponential will restrict the integral to small $z$, i.e. the threshold
region. As a result the ratio will behave as a power law. All together,
\begin{equation}
\label{etaoversxxx}
  \frac\eta s = N^2\ZZ^{9/4}\sqrt{\c A_N}\, \left\{
    \begin{array}[c]{ll}
      \rule[-1.5em]{0em}{1em}\displaystyle \frac{512}{12^{5/4}\,\pi}\,
      \left(\frac{2T}{\ZZ^{2/5}\sigma}\right)^{5/2}\,
      \qquad & \mathrm{if}\; T\ll\ZZ^{2/5}\sigma \cr  
      \displaystyle \frac{2}{15}\, \sqrt{\frac{5+2\sqrt{5}}{10\, \pi\,
          \sqrt{3}}} \, \approx\, 0.056 \qquad & \mathrm{if}\;
      T\gg\ZZ^{2/5}\sigma 
    \end{array}\right.
\end{equation}
Here high temperature of course is still limited to the confinemened phase
since we are using the strong coupling expansion. In particular we are not
expected to reach the deconfined phase
and so the expected increase of $\eta/s$ for increasing $T$ above $T_c$ can
not be seen here.

Let us note that the result (\ref{etaoversxxx}) scales as $N^2$ in accordance with the
expectation that for $T < T_c$ the entropy density scales as $N^0$ and
the shear viscosity as $N^2$.
Also, in the high temperature limit the $T$ dependence drops
out.  It may be in connection to the fact that at high
temperature the relevant part of the spectrum is the peak region, which is
pretty similar to a real quasi particle peak. The physics of the low
temperature asymptotics is rather different -- there the relevant part is the
threshold regime with continuous number of excitations.

The most robust part of our prediction is the functional dependence of the
$\eta/s$ ratio on the temperature. Unfortunately both in the absolute magnitude
of the ratio and in the temperature scale, the lattice spacing dependent $\ZZ$
quantity is still present. It is, in fact, a common feature of the strong
coupling expansion results, as it was discussed in the introduction. For
the physical result we may use the form $\frac{\eta}s = K f(T/T_0)$ where $K$
and $T_0$ are the ``renormalized'' value of the prefactor and temperature
scale in \eqref{etaperes_sce}.

To assess $K$ we may argue that at small $\ZZ$ (large lattice spacing) we are
safely in the applicability range of the strong coupling expansion, but far
from the continuum limit. On the other hand at large $\ZZ$ (small lattice
spacing) we are near the continuum limit, but far from the applicability range
of the strong coupling expansion. As a best compromise we may take the
prefactor at $\ZZ\approx1$, which suggests $K=1$. 

It is even harder to assess the value of $T_0$, which is approximately the
``turning point'' of the $\eta/s$ ratio towards the low temperature
asymptotics. We may argue that it is the highest temperature end of our
calculation where we are near the continuum limit and the strong coupling
expansion applicability range. It is then natural to expect that the $T_0$
scale is determined at $\ZZ\approx1$ and $\sigma$ is the glueball mass near
the phase transition. This consideration leads to,
\begin{equation}
  \label{etaperes_phys}
  \frac\eta s = \left\{
    \begin{array}[c]{ll}
      4\quad &\mathrm{for\ N = 2}\cr
      N^2/2\quad&\mathrm{for\ N > 2}\cr
    \end{array}\right\}\times f(T/M_{\mathrm{glueball}}(T_c)).
\end{equation}
As a consequence, from (\ref{etaoversxxx}) and (\ref{etaperes_phys}), 
we estimate the shear viscosity to entropy density ratio at $T_c$ to be
\begin{equation}
  \frac\eta s = \left\{
    \begin{array}[c]{ll}
      0.22\quad &\mathrm{for\ N = 2}\cr
      0.028 N^2\quad&\mathrm{for\ N > 2}\cr
    \end{array}\right..
\end{equation}
This is the second main result of our paper.

\section{Discussion}
\label{sec:discussion}

In this paper we discussed the shear viscosity to entropy density ratio in
pure $SU(N)$ gauge theories. The strong coupling expansion of the Euclidean lattice
formulation of the theory was used.
This expansion
is known to be convergent and it yields physical results for a number
of low energy observables \cite{Montvay:1994cy}, on the other hand its
renormalization is not a fully clarified topic. 
This is reflected in our result too, since both in the magnitude of the
$\eta/s$ ratio and also in the temperature scale a lattice spacing dependent
factor was present. For a physical interpretation we used the robust
(i.e. lattice spacing independent) prediction of the calculation, which was the
functional form of the dependence of $\eta/s$ on the temperature and assessed
the scale factors at ``half way'' between the strict strong coupling
applicability range and the continuum limit. As a result we estimate the
$\eta/s$ ratio for gauge group $SU(N)$ at $T \approx T_c$ as 
$0.22$ for $N = 2$ and $0.028 N^2$ for $N > 2$ which is consistent with the
conjectured lower bound $1/4\pi$ inspired by AdS/CFT ideas.

The temperature dependence of $\eta/s$ consists of two well separated
regimes. At higher temperatures (but still in the low temperature phase of the
gauge theory) we see a very flat function, slightly rising towards smaller
temperatures. This is the quasi particle regime, where we are in good agreement
with the AdS/CFT prediction. At low temperature we arrive at a different
regime, where the near-threshold part of the spectral function dominates, and
so it is rather far from a quasi particle picture. The ``continuum number'' of
peaks, forming a continuum, allow to compare to the constructions given in
\cite{Cohen:2007qr, Cohen:2007zc}. In particular we observe a vanishing
$\eta/s$ ratio near zero temperature.

\section*{Acknowledgements}

DN would like to acknowledge helpful discussions with Julius Kuti and Barak Bringoltz. In
addition we are grateful to Owe Philipsen for email correspondence.
The work of DN was supported by the DOE under grants DOE-FG03-97ER40546 and DE-FG02-97ER25308.
AJ acknowledges support from the Humboldt Foundation and the Hungarian Science
Fund (OTKA) under grant K68108.

\end{document}